\PassOptionsToPackage{dvipsnames}{xcolor}
\documentclass[10pt,showpacs,letterpaper,aps,nofootinbib,prx, twocolumn,superscriptaddress]{revtex4-2} 

\usepackage{tikz}
\usetikzlibrary{patterns,arrows,calc,decorations.pathmorphing,positioning}

\usepackage{amsmath,amssymb,amsfonts,amsthm,gensymb,siunitx,bm,physics,mathtools,graphicx,tabularx,tabulary,booktabs,multirow,array,float,subcaption,csquotes,comment,bbold,syntonly,textcomp,centernot}
\usepackage[title]{appendix}
\usepackage[format=plain,labelfont=it,textfont=it]{caption}
\usepackage{chngcntr}
\usepackage{color}
	\definecolor{mygray}{RGB}{222, 222, 222}
\usepackage{enumitem}
	\setlist{nosep}
\usepackage{epstopdf}
\usepackage[none]{hyphenat}
\usepackage[version=4]{mhchem}
\usepackage[indent,parfill]{parskip}		
\usepackage[nottoc,numbib]{tocbibind}
	\counterwithout{figure}{section}
\usepackage{xparse}				
\usepackage{dsfont}
\usepackage[colorlinks=false]{hyperref} 
\usepackage[capitalize]{cleveref}
\usepackage{soul}
\usepackage{tabstackengine}
\setstackEOL{\cr}

\usepackage{thm-restate}
\newtheorem{theorem}{Theorem}[]

\newtheorem{lemma}[theorem]{Lemma}
\newtheorem{definition}[theorem]{Definition}

\newcommand{\beq}{\begin{equation}}
\newcommand{\eeq}{\end{equation}}
\newcommand{\bea}{\begin{align}}
\newcommand{\eea}{\end{align}}

	\DeclarePairedDelimiterX{\Set}[1]{\{}{\}}{%
		
		#1
	}
	\makeatletter
		\let\oldSet\Set
		\def\Set{\@ifstar{\oldSet}{\oldSet*}}
	\makeatother

	\DeclarePairedDelimiterX{\Family}[1]{(}{)}{%
		
		#1
	}
	\makeatletter
		\let\oldFamily\Family
		\def\Family{\@ifstar{\oldFamily}{\oldFamily*}}

	\newsavebox{\numbox}%
	\newsavebox{\slashbox}%
	\newsavebox{\denbox}%
	\newlength{\slashlength}%
	\newlength{\faktorscale}%
	\DeclareDocumentCommand{\newfaktor}{m O{0.35} m O{-0.35}}{
		\savebox{\numbox}{\ensuremath{#1}}
		\savebox{\slashbox}{\ensuremath{\diagup}}
		\savebox{\denbox}{\ensuremath{#3}}
		\setlength{\faktorscale}{0.5\ht\numbox+0.5\ht\denbox}%
		\setlength{\slashlength}{2pt+0.8\faktorscale+#2\faktorscale-#4\faktorscale}%
		\raisebox{#2\ht\slashbox}{\usebox{\numbox}}
		\mkern-2mu%
		\rotatebox{-30}{\rule[#4\ht\denbox]{0.4pt}{\slashlength}}
		\mkern9mu%
		\hspace{-0.44\slashlength}%
		\raisebox{#4\ht\denbox}{\usebox{\denbox}}
	}
	\DeclareDocumentCommand{\linefaktor}{m O{0.08} m O{-0.08}}{
		\savebox{\numbox}{\ensuremath{#1}}
		\savebox{\slashbox}{\ensuremath{\diagup}}
		\savebox{\denbox}{\ensuremath{#3}}
		\setlength{\faktorscale}{0.5\ht\numbox+0.5\ht\denbox}%
		\setlength{\slashlength}{0.2\faktorscale+0.8\baselineskip}%
		\raisebox{#2\ht\slashbox}{\usebox{\numbox}}
		\mkern-1mu%
		\raisebox{-0.8pt}{%
			\rotatebox{-30}{\rule[#4\ht\denbox]{0.4pt}{\slashlength}} 
		}%
		\mkern-1mu%
		\hspace{-0.25\slashlength}%
		\raisebox{#4\ht\denbox}{\usebox{\denbox}}
	}

\definecolor{js}{rgb}{0.578125,0.23828125,0.9609375}

\newcommand{\blk}{\color{black}}

\begin{document}

\title{Symmetry-induced failures of tomographic locality:\\Constructing foil theories by twirling}

\author{Daniel Centeno}
\affiliation{Perimeter Institute for Theoretical Physics, Waterloo, Ontario, Canada, N2L 2Y5}
\affiliation{Department of Physics and Astronomy, University of Waterloo, Waterloo, Ontario, Canada, N2L 3G1}
\author{Marco Erba}
\affiliation{International Centre for Theory of Quantum Technologies, University of Gda\'nsk, 80-309 Gda\'nsk, Poland}
\author{Thomas D. Galley}
\affiliation{Institute for Quantum Optics and Quantum Information, Austrian Academy of Sciences, Boltzmanngasse 3, A-1090 Vienna,Austria}
\affiliation{Vienna Center for Quantum Science and Technology (VCQ),  Faculty of Physics, University of Vienna, Vienna, Austria}
\author{David Schmid}
\affiliation{International Centre for Theory of Quantum Technologies, University of Gda\'nsk, 80-309 Gda\'nsk, Poland}
\author{John H.~Selby}
\affiliation{International Centre for Theory of Quantum Technologies, University of Gda\'nsk, 80-309 Gda\'nsk, Poland}
\affiliation{Theoretical Sciences Visiting Program, Okinawa Institute of Science and Technology Graduate University, Onna, 904-0495, Japan}
\author{Robert W. Spekkens}
\affiliation{Perimeter Institute for Theoretical Physics, Waterloo, Ontario, Canada, N2L 2Y5}
\affiliation{Department of Physics and Astronomy, University of Waterloo, Waterloo, Ontario, Canada, N2L 3G1}
\author{Sina Soltani}
\affiliation{International Centre for Theory of Quantum Technologies, University of Gda\'nsk, 80-309 Gda\'nsk, Poland}
\author{Jacopo Surace}
\affiliation{Perimeter Institute for Theoretical Physics, Waterloo, Ontario, Canada, N2L 2Y5}
\author{Alex Wilce}
\affiliation{Department of Mathematical Sciences, Susquehanna University, Selinsgrove, PA 17870, United States}
\author{Y{\`i}l{\`e} Y{\=\i}ng}
\thanks{The author order is alphabetical. \newline Correspondence email: \href{yile.ying@gmail.com}{yile.ying@gmail.com}}
\affiliation{Perimeter Institute for Theoretical Physics, Waterloo, Ontario, Canada, N2L 2Y5}
\affiliation{Department of Physics and Astronomy, University of Waterloo, Waterloo, Ontario, Canada, N2L 3G1}

\begin{abstract}
    Tomographic locality is a principle commonly used in the program of finding axioms that pick out quantum theory within the landscape of possible theories.  
    The principle asserts the sufficiency of local measurements for achieving a tomographic characterization of any bipartite state. In this work, we explore the meaning of the principle of tomographic locality by developing a simple scheme for generating a wide variety of theories that violate the principle. In this scheme, one starts with a tomographically local theory---which can be classical, quantum or post-quantum---and a physical symmetry, and one restricts the processes in the theory to all and only those that are covariant with respect to the collective action of that symmetry. We refer to the resulting theories as {\em twirled worlds}.  We show that failures of tomographic locality are ubiquitous in twirled worlds. From the possibility of such failures in {\em classical} twirled worlds, we argue that the failure of tomographic locality (i.e., tomographic {\em non}locality) does not imply ontological holism.  Our results also demonstrate the need for researchers seeking to axiomatize quantum theory to take a stand on the question of whether there are superselection rules that have a fundamental status. 
\end{abstract}

\maketitle

To better understand what is distinctive about quantum theory in terms of its operational predictions, it is useful to conceptualize it as a point in a space of possible theories. 
Many frameworks have been proposed for describing such theories, with the framework of Generalized Probabilistic Theories (GPTs)~\cite{hardy2001quantum,barrett2007information,Chiribella2010purification,muller2021probabilistic,Janotta_2014,Plavala_2023} being a particularly popular choice.  
The project of articulating axioms relative to which quantum theory, considered as a GPT\footnote{In this article, the term ``quantum theory'' will always refer to quantum theory considered as a GPT.}, is uniquely picked out from the landscape is known as the {\em quantum reconstruction program}~\cite{muller2021probabilistic,hardy2013reconstructing,Wilce_2019}.
{\em Tomographic locality}\footnote{The principle of tomographic locality has  various names in the literature, e.g., {\em local discriminability} and {\em local tomography}.  The fact that it is satisfied in quantum theory was noted in several works prior to the reconstruction program, e.g., in Ref.~\cite{Mermin_1998,Wooter_1990,arakiCharacterization1980,bergiaActual1980}. } is a principle appearing in many such axiomatic schemes~\cite{hardy2001quantum,barrett2007information,Chiribella2010purification,zaopo2012information,hardy2013reconstructing,Masanes_2013,Muller_2015,muller2021probabilistic,Hohn2017,wilce2009halfaxiomsfinitedimensional,dakic2009quantumtheory,selby2021reconstructing}.  It asserts that for the purpose of achieving a tomographic characterization of a bipartite state (i.e., inferring the state from the measurement statistics it induces), it is sufficient to use only local measurements.  That is, it is sufficient to use measurements carried out separately on each individual system (i.e., separable measurements), rather than requiring measurements that must be implemented jointly on the pair of systems (i.e., entangled measurements).

The best way to understand the content of an axiom proposed in the quantum reconstruction program is to study theories in which it fails.  In such an approach, alternatives to quantum theory are studied not because they are thought to be empirical competitors, but because they serve as {\em foils}, helping to highlight what is truly distinctive about quantum theory by contrasting with it~\cite{Chiribella2016}.  
The main example cited in the literature of a foil theory exhibiting a failure of tomographic locality is quantum theory over the field of reals, or {\em Real Quantum Theory}. 
Other examples are
quaternionic quantum theory~\cite{hardy2001quantum}, fermionic quantum theory~\cite{darianoFermionic2014,Dariano2014feynmanproblem}, exotic variants of classical theories~\cite{d2020classicality,scandolo2019information,Chiribella2024} and of quantum theory~\cite{erba2024compositionrulequantumsystems}, and a post-quantum theory constructed from Euclidean Jordan algebras~\cite{barnum2020composites}.
More examples would clearly be useful.  
In this article, we present a scheme for generating a wide variety of distinct foil theories that exhibit a failure of tomographic locality.  We do so by starting with a tomographically local GPT and imposing a symmetry, thereby defining a new theory that we refer to as a \emph{twirled world}. 

We demonstrate this scheme by imposing symmetry on a variety of different GPTs, including quantum, classical, and post-quantum theories, and using a variety of different symmetries, 
and we prove that failures of tomographic locality {\em always} arise from it when there are states of the GPT that transform nontrivially under the collective action of that symmetry.

{\bf Defining twirled worlds---}
We will describe the basic formalism of twirled worlds using quantum theory. The generalization to the case of arbitrary GPTs, 
which will be important later, is given in the  appendix.  

 For a given symmetry, such as spatial translations, rotations, time translations, as well as reversible transformations of internal degrees of freedom, it is presumed that the appropriate representation of the symmetry group on a given system is part of the content of the theory under consideration. Quantum theory, for instance, specifies a particular unitary representation of the group of spatial rotations for every type of system---integer spin, half-integer spin, particle, etcetera. The representation of such a symmetry on a composite system is simply the collective representation, where the same group element is applied to each component. 

Specifically, consider a quantum system $S$ composed of $n$ subsystems, denoted $S_1$, $S_2$, $\dots$, $S_n$. Consider a symmetry that is described by group $G$ and by a projective unitary representation\footnote{Such a representation is in terms of unitaries $U_g$ such that $U_g U_{g'} = e^{i\theta(g,g')}U_{g\cdot g'}$ where $\theta(g,g')\in [0,2\pi)$.},  $\{U^{S_i}_g\}_{g\in G}$ on the Hilbert space of subsystem $S_i$. 
Let $\{ \mathcal{U}^{S_i}_g \}_{g\in G}$ 
denote the associated {\em superoperator} representation of this symmetry group\footnote{ Note that the phases $e^{i\theta(g,g')}$ from the projective representation cancel out when we go to the superoperator representation, i.e., the superoperators satisfy $\mathcal{U}_g\circ\mathcal{U}_{g'}=\mathcal{U}_{g\cdot g'}$. \blk }, that is, the representation on the space of Hermitian operators of system $S_i$ such that  for any quantum state (i.e., density operator) $\rho^{S_i}$, the action associated to $g\in G$ is
\begin{align}
\mathcal{U}^{S_i}_g[\rho^{S_i}] := U^{S_i}_g \rho^{S_i} {U^{S_i}_g}^\dagger = U^{S_i}_g \rho^{S_i} U^{S_i}_{g^{-1}}.
\end{align}
Similarly, for any effect  $E^{S_i}$ (i.e., any element of a positive operator-valued measure on system $S_i$), the action associated to $g\in G$ is also
\begin{align}
\mathcal{U}^{S_i}_g [E^{S_i}] := {U^{S_i}_g} E^{S_i} {U^{S_i}_g}^\dagger = U^{S_i}_{g} E^{S_i} U^{S_i}_{g^{-1}}.
\end{align} 

The superoperator representation for a 
composite system $S=S_1 S_2 \dots S_n$ is the collective representation\footnote{For different subsystems of $S$, e.g., $S_1$ and $S_2$, the representations ${\cal U}^{S_1}$ and  ${\cal U}^{S_2}$ may be different. For example, the representation for spatial rotations depends on the spin of the system. Note, however, that for the examples we consider here, composite systems are made up of systems that are all of the same type.} 
\begin{align}
\label{eq:collective}
\{\mathcal{U}^{S}_g = \mathcal{U}^{S_1}_g \otimes \mathcal{U}^{S_2}_g\otimes \dots  \otimes \mathcal{U}^{S_n}_g\}_{g\in G}.
\end{align} 

We say that a Hermitian operator $O^{S}$ (state or effect) is $(G,\mathcal{U})$-invariant if and only if 
\begin{align}
  \mathcal{U}^{S}_g [O^{S} ] =O^{S} \ \ \forall g\in G.
\end{align}
Details about the characterization of such operators in terms of the representation theory of $G$ are provided in the appendix. 

A quantum operation from $S$ to $S'$ is represented in quantum theory by a completely positive trace-nonincreasing map  $\mathcal{E}^{S'|S}$.   Such a map is said to be $(G,\mathcal{U})$-covariant if  and only if 
\begin{align}
  \mathcal{E}^{S'|S} \circ  \mathcal{U}^{S}_g =  \mathcal{U}^{S'}_g\circ \mathcal{E}^{S'|S} \ \ \forall g\in G. 
\end{align}

When $G$ is compact, there is a natural linear map $\mathcal G$ taking arbitrary operators to $(G,\mathcal{U})$-invariant operators. This is defined by group averaging:
\begin{align}
\mathcal{G}(O^S):= \int_G {\rm d}g\ \mathcal{U}^S_g [O^S]
\end{align}
where ${\rm d}g$ is the Haar measure on $G$. Following common usage in quantum information theory, we call $\mathcal G$ the {\em $(G,\mathcal{U})$-twirling map}.
When $G$ is a discrete group, the integral is replaced by a sum.
Similarly, one can define a $(G,\mathcal{U})$-twirling {\em supermap},  i.e., $\frac{1}{|G|}\sum_{g\in G} \mathcal{U}_g \circ \cdot \circ  \mathcal{U}_g^{\dag}$, that takes all quantum operations onto the $(G,\mathcal{U})$-covariant operations.
  
The twirling map is idempotent, and sends states to states and effects to effects. Thus, a state or effect is invariant if and only if it arises from twirling a state or effect. 
A {\em $(G,\mathcal{U})$-twirled world} for a set of quantum systems is a subtheory of quantum theory on those systems where the states and effects are restricted to all and only those that are $(G,\mathcal{U})$-invariant and where the quantum operations are restricted to all and only those that are $(G,\mathcal{U})$-covariant.\footnote{A fully compositional approach to quantum theory with covariant processes can be found in Def.~2.15 of Ref.~\cite{selby2022time}.}

It is well-known that in the case of quantum theory, a restriction to symmetric states, measurements, and operations  can be viewed as a superselection rule~\cite{Bartlett_2007}, so that a twirled quantum world is an instance of 
a quantum world with a superselection rule.

{\em Phase-shift-twirled bosonic world---}
Consider, as the system of interest, a collection of modes for some type of bosonic field, and as the symmetry, the phase-shifts for the phase conjugate to number. Here the symmetry group is U(1) and the relevant unitary representation thereof is $\{ e^{-i \phi \hat{N}} \}_{\phi \in U(1)}$, where $\hat{N}$ is the total number operator. 

The corresponding superoperator representation is $\mathcal{U}_\phi(\cdot) = e^{-i \phi \hat{N}}(\cdot) e^{i \phi \hat{N}}$, which implies the possibility of a nontrivial phase shift only for operators that have nontrivial off-diagonal components relative to the eigenspaces of $\hat{N}$~\cite{Marvian2014modes}.
Hence, the invariant operators are those that are block-diagonal relative to these eigenspaces. 
We refer to the theory one obtains by imposing this symmetry as the {\em phase-shift-twirled bosonic world}. It can also be conceptualized as the quantum theory of bosonic modes after imposing a U(1)-superselection rule, where the superselection sectors are the eigenspaces of total boson number~\cite{Bartlett_2007}. 
We will now show that this twirled world fails tomographic locality. 

For a single bosonic mode, (i.e., a unipartite system), the eigenspaces of the number operator are all one-dimensional.  Specifically, for a given eigenvalue $n \in \mathbb{Z}_+$, the eigenspace is the span of the single state $\ket{n}$.
 Thus, the single-mode invariant states and effects are all and only those that are diagonal in the number eigenbasis. 
 
 For a pair of bosonic modes, the total number eigenspaces are no longer all one-dimensional since there are $n+1$ ways to distribute $n$ excitations
   between a pair of modes.  For $n=1$, for example, one can put the excitation in either mode, so that the associated total number eigenspace is the span of $\ket{0}\ket{1}$ and $\ket{1}\ket{0}$. Since the invariant states merely need to be block-diagonal with respect to the total number eigenspaces, they can constitute nontrivial superpositions of the states within such a space. For example, all states of the form $(1/\sqrt{2})\left(|0\rangle|1\rangle + e^{i\theta} |1\rangle|0\rangle\right)$ for different values of $\theta$ are invariant under collective phase-shifts. However, the only invariant measurement on each mode is a measurement of the local number operator, so the statistics obtained from local invariant measurements reveal nothing about the value of $\theta$.\footnote{This fact was leveraged in Ref.~\cite{verstraete2003quantum} to show the possibility of quantum data hiding when the parties lack a shared reference frame.}  It follows that the phase-shift-twirled bosonic world exhibits a failure of tomographic locality.

{\em Rotation-twirled spinor world---} 
Consider, as the system of interest, a collection of spinors (where by ``spinor'' we mean the spin degree of freedom of a spin-1/2 particle) and as the symmetry, the spatial rotations. Here, the relevant unitary representation is the spinor representation, SU(2), where the group actions are represented by Wigner rotation operators. We refer to the theory one obtains by imposing this symmetry as the {\em rotation-twirled spinor world}. It can also be conceptualized as the quantum theory of spinors after imposing an SU(2)-superselection rule~\cite{Bartlett_2007}.

A single spinor carries a spin-$1/2$ irreducible representation of SU(2); thus, in the rotation-twirled spinor world, the only invariant operator is the identity operator up to a constant.
Thus, there are no informative measurements on a single spinor in the twirled world. For two spinors, the joint Hilbert space decomposes into a direct sum of the singlet subspace and the triplet subspace, each of which carries an irreducible representation of SU(2) and has no multiplicity. Thus, any bipartite state that is a convex mixture of the singlet state and the projector onto the triplet subspace is rotationally invariant. However, this convex weight cannot be estimated using local measurements, given that the latter are trivial. That is,  all invariant bipartite states look the same under local measurements. Therefore, the rotation-twirled spinor world exhibits a failure of tomographic locality.\footnote{See appendix for an alternative proof of this failure 
based on parameter counting. }

{\bf The status of superselection rules impacts the status of tomographic locality---} 
There has been a long debate on whether various superselection rules (SSRs)  are (i) fundamental, so that they should be considered as additional axioms of quantum theory~\cite{wick1997intrinsic,giulini2007superselection}, or (ii) arising merely from technological limitations in certain circumstances and therefore in-principle circumventable. See, e.g.,~\cite{wick1997intrinsic,aharonov1967charge,bartlett2006dialogue}.

Support for option (ii) comes from the fact that many of the SSRs that were the focus of this debate (such as the U(1) SSR considered above) can arise from twirling, i.e., they are \emph{symmetry-induced} SSRs~\cite{Bartlett_2007}.
The restriction to the symmetric states, effects and operations can be understood as arising from the lack of an appropriate reference frame and consequently
 can be {\em lifted} by considering the relational degrees of freedom relative to such a reference frame~\cite{aharonov1967charge, kitaev2004superselection,bartlett2006dialogue,Bartlett_2007,dowling2006observing}. 
However, the question about whether {\em all} symmetry-induced SSRs can be lifted remains open.  In particular, no prescription is known for lifting the univalence SSR~\cite{giulini2007superselection,dowling2006observing} (which forbids a superposition of even and odd fermionic numbers), suggesting that it might be fundamental, and hence that option (i) holds.

The following example shows how the univalence SSR can arise from imposing a parity twirling on fermionic quantum theory, and how the parity-twirled fermionic world fails tomographic locality.  

{\em Parity-twirled fermionic world---} 
We consider the parity representation of $\mathbb{Z}_2$, namely, $\{\mathbb{1}, \hat{\pi}\}$ where $\hat{\pi}\coloneq (-1)^{\hat{N}}$ is the parity operator ($\hat{N}$ is the number operator). For $n$ fermionic modes, the eigenspaces of $\hat{\pi}^{\otimes n}$ (associated to eigenvalues +1 and -1) are the subspaces of even and odd total fermion number. 
For a single fermionic mode, therefore, the only invariant operators are those that are diagonal in the number eigenstates  $\ketbra{0}$ and $\ketbra{1}$. Consequently, the only invariant local measurement is a measurement of the local number. 
For a {\em pair} of fermionic modes, the invariant operators are those that are block-diagonal in
the parity eigenspaces, which are  span$\{\ket{00},\ket{11}\}$ (even parity) and span$\{\ket{01},\ket{10}\}$ (odd parity).  Thus, states on the pair of modes of the form $(1/\sqrt{2})(\ket{01}+e^{i\theta}\ket{10})$ are invariant under parity, hence preparable in the parity-twirled fermionic world.  Clearly, however, no information about the phase $\theta$ can be determined from  measurements of local number. As such, the parity-twirled fermionic world fails to be tomographically local. Our parity-twirled fermionic world is equivalent to the ``fermionic quantum theory'' defined in Ref.~\cite{darianoFermionic2014}, where the failure of tomographic locality for this theory was first noted. We note that this is an example of two classical bits composing to a pair of superselected qubits, showing the central importance of the composition rule when defining a theory.

{\em Implications---}
Because the debate on whether there are any SSRs that are fundamental has concerned SSRs that arise from twirling, and because twirled worlds always exhibit a failure of tomographic locality (see Theorem~\ref{thm:ubiquity} below), 
the status of symmetry-induced SSRs is tethered to the status of tomographic locality.
\footnote{Note that there are foil theories described as ``quantum theory with a SSR'' that \emph{satisfy} tomographic locality~\cite{barnum2014Local,wilce2018Royal,selby2021reconstructing,nakahira2020Derivation,westerbaan2022computer}. However, these SSRs do not arise from twirling and consequently are not the type that has appeared in the debate on whether some SSRs are fundamental. } If one takes the view that some of these SSRs are fundamental  (option (i) above), then one must grant that not all types of quantum systems satisfy tomographic locality, and therefore if one pursues the quantum reconstruction program, one must take on the task of articulating the scope of the principle's applicability. If, on the other hand, one wants to avoid this conclusion and instead endorse the view that no symmetry-induced SSR is fundamental (option (ii) above), one must justify the claim that all symmetry-induced SSRs can be lifted in principle, since this has not been established in prior work.

More generally, to the extent that there is still debate about whether certain SSRs are fundamental, the question of which of the two theories---quantum theory without any SSRs or quantum theory with some  SSRs---is the empirically most successful one (hence the one that should be the target of reconstruction efforts) remains unresolved. 
That is, certain twirled quantum worlds become contenders for the target of reconstruction efforts, 
rather than mere foils.

{\bf A failure of tomographic locality does not imply ontological holism---}
A failure of tomographic locality is often described as a type of holism~\cite{darianoFermionic2014,Hardy2011Limited}. But does it imply holism in the conventional ontological sense? More precisely, if a theory fails to be tomographically local, does it imply that any ontological account of the predictions of this theory must involve a failure of ontological reductionism?  We here consider this question under the most conservative notion of an ontological account, namely, the one provided by the framework of ontological models~\cite{Harrigan_2010}, where ontological reductionism is defined by the condition that the ontic state space of a composite system {\em factorizes} into a Cartesian product of the ontic state spaces of the component subsystems.\footnote{This assumption is also sometimes termed `ontic separability' or `kinematic locality'.}  
We answer the question in the negative by appealing to examples of twirled {\em classical} worlds. \blk

{\em Bit-flip-twirled cbit world---}
Consider a world with systems that are classical bits (cbits for short). We denote the ontic state space of a classical bit by $\Lambda = \{0,1\}$. The ontic state space of two cbits, $A$ and $B$, is given by the Cartesian product of their ontic state spaces, so that $\Lambda_{AB} = \Lambda_A \times \Lambda_B =\{(0,0),(0,1),(1,0),(1,1)\}$. 
The analog of a quantum/GPT state in a classical world is a {\em statistical} state, i.e., a probability distribution over the ontic state space, and effects are response functions over the ontic state space. Consequently, when we consider a twirled version of this theory, we are considering a restriction on the statistical states and response functions, without any change to the ontic state space on which these are defined. We consider the symmetry associated to the $\mathbb{Z}_2$ group consisting of identity and bit-flip. 
For a single cbit, the only invariant distribution is the uniform one. 
There is also only a single invariant response function, assigning probability 1 to both values of the cbit, which is simply the unit effect in this theory. 

For a pair of cbits, the only invariant distributions are those that are uniform over bitstrings having a fixed parity, but with an arbitrary distribution over the different parities, i.e., \begin{equation}
P_{AB} = w ( \tfrac{1}{2}[00]+\tfrac{1}{2}[11])+(1-w)( \tfrac{1}{2}[01]+\tfrac{1}{2}[10])    
\end{equation} 
where $[ab]$ denotes a point distribution where system $A$ takes value $a$, system $B$ takes value $b$, and where $w\in[0,1]$. 
Since there are no nontrivial invariant measurements on a single cbit, 
the convex weight $w$ cannot be characterized by such measurements, and consequently we have a failure of tomographic locality.

Note that the failure of tomographic locality does not depend on the classical system being discrete, as the following example shows.

{\em Rotation-twirled pointer world---}
In this world, each system is a pointer in 3-space, so that the ontic state space corresponds to the set of possible directions, i.e., the sphere. The ontic state space of multiple pointers is simply the Cartesian product of these spheres. Each system supports a representation of SO(3) in terms of rotations of the pointer; for composite systems, the relevant symmetry is collective rotations.   

For a single such system, the only rotationally invariant probability distribution is the uniform distribution on the sphere, and the only rotationally invariant response function is the unit effect. Again, therefore, the theory contains no informative measurements on a single system.

For the bipartite system, there are many nontrivial rotationally invariant distributions, namely, distributions whose marginals are uniform, but for which the {\em relative orientation} of the arrows is not uniformly distributed. 
The key is that the relative orientation does not change under a collective rotation.  
Because this relative angle cannot be estimated by local invariant measurements (since these are uninformative), it follows that tomographic locality fails in this theory.

{\em Implications---}
In these two twirled worlds, there is a failure of tomographic locality even though ontological reductionism holds. For these examples, the classical world from which the twirled world is obtained as the image of a twirling operation is {\em itself} a reductionist ontological theory, and therefore it provides a reductionist ontological model of the predictions of the twirled world.\footnote{As another example, the foil theory known as bilocal classical theory~\cite{d2020classicality}, which exhibits a failure of tomographic locality,  can also be shown~\cite{soltani2024bilocal} to admit of a reductionist ontological model.
}

It follows that tomographic locality cannot be motivated by the intuition that ontological holism is unnatural, undermining what might previously have been taken to be a strong reason in favor of adopting the principle.

{\bf Symmetry-induced failures of tomographic locality are ubiquitous---}\label{ubiquitous}
Symmetry-induced failures of tomographic locality occur in arbitrary theories, not merely quantum or classical ones.
In the appendix, we generalize the definition of twirled worlds to arbitrary GPTs.
We then show (in the appendix) that symmetry-induced failures of tomographic locality are ubiquitous in the following sense:
\begin{restatable}{theorem}{ubiquity}\label{thm:ubiquity}
For any GPT $\mathds{T}$ (not necessarily classical or quantum) and for any physical symmetry group $G$ with an action $\mathcal{V}$ that is nontrivial on some state(s)
within $\mathds{T}$, the corresponding $(G,\mathcal{V})$-twirled $\mathds{T}$-world exhibits a failure of tomographic locality.
\end{restatable}

Throughout this article, we have provided many concrete examples of symmetry-induced failures of tomographic locality starting from classical and quantum theories.  One can just as easily devise concrete examples starting from a post-quantum theory.  In the appendix, we do so for the paradigm example of a post-quantum theory, namely, boxworld~\cite{barrett2007information}.

{\bf Discussion---}
The motivation for studying foils to quantum theory is the idea that it is only by studying the contrast class of a phenomenon that one can fully understand it.  One might refer to this approach as the {\em methodology of foil theories}. 
 We have here applied this methodology to the study of tomographic locality.\footnote{In forthcoming work~\cite{ying2024twirling2}, we apply this methodology to the study of \emph{tomographic $N$-locality}~\cite{Hardy2011Limited}, of which tomographic locality is the special case when $N=1$. We find that certain twirled worlds satisfy tomographic $N$-locality for some $N > 1$.}  Specifically, we have shown a scheme for generating a number of foil theories that exhibit a failure of tomographic locality: one starts with a physical symmetry and a classical, quantum, or post-quantum theory, and one restricts the processes in the theory to those that are covariant with respect to that symmetry.  These theories, which we refer to as {\em twirled worlds}, shed light on the interpretation of the principle of tomographic locality, in particular, that its failure need not imply ontological holism. They also show that reconstructionists need to take a stand on the question of whether there are superselection rules that have a fundamental status in quantum theory.

For many other phenomena related to quantum entanglement theory and, more generally, to the manner in which quantum systems compose, it is known that imposing symmetry can change the nature of the phenomenon.
Several works ~\cite{bartlett2005mixed,schuch2004quantum,vaccaro2008tradeoff,jones2006entanglement,wiseman2004ferreting,bartlett2003entanglement,Bartlett_2007} have highlighted 
aspects of entanglement theory that change significantly in the presence of a superselection rule.  For instance, the phenomenon of {\em bound entanglement} arises only for mixed states in standard quantum theory, while it can arise for {\em pure states} when
a superselection rule is imposed~\cite{bartlett2005mixed}.  Similarly, the monogamy of pure entanglement, i.e.,  that if two systems are correlated and in a convexly extremal (i.e., pure) state, then neither can be correlated with any other system, generally fails under a superselection rule~\cite{darianoFermionic2014,Dariano2014feynmanproblem}.
One can understand such works through the lens of the methodology of foil theories, and see what lessons they hold for the interpretation of various phenomena in entanglement theory.  Prior works in this space have not availed themselves of the full power of this methodology insofar as they confined attention to foil theories that were subtheories of quantum theory. 
  Many insights might require consideration of foil theories that are restrictions on classical or post-quantum theories, as was the case in the present work.

{\bf Acknowledgements---} For useful discussions, we thank Nick Ormrod, Yujie Zhang, Jinmin Yi, Ruizhi Liu, and Jan G{\l}owacki.
YY, DC, JS and RWS are supported by Perimeter Institute for Theoretical Physics. 
AW was on sabbatical at Perimeter Institute while some of this research was being undertaken and would like to thank the quantum foundations group there for their hospitality during his visit. Research at Perimeter Institute is supported in part by the Government of Canada through the Department of Innovation, Science and Economic Development and by the Province of Ontario through the Ministry of Colleges and Universities. 
YY is also supported by the Natural Sciences and Engineering Research Council of Canada (Grant No. RGPIN-2024-04419). 
SS, DS, ME, and JHS were supported by the National Science Centre, Poland (Opus project, Categorical Foundations of the Non-Classicality of Nature, project no. 2021/41/B/ST2/03149). JHS conducted part of this research while visiting the
Okinawa Institute of Science and Technology (OIST) through the Theoretical Sciences Visiting
Program (TSVP).

\bibliography{reference.bib}

\appendix

\section{Symmetry and twirling in quantum theory}
\label{app_SSR}

Here we list some of the key results on 
the connection between superselection rules and $(G,\mathcal{U})$-twirling maps. For more details on this connection, we refer the reader to Ref.~\cite[II.C]{Bartlett_2007}.

Given a unitary symmetry described by a group $G$  (a compact Lie group, possibly finite),  the Hilbert space of a system can be decomposed into a direct sum of sectors $\mathcal{H}_q$, as 
\begin{align}
	\mathcal{H}=\bigoplus_q \mathcal{H}_q,
\end{align}
where each sector carries an inequivalent representation of $G$ labeled by $q$. 
Furthermore, each sector can be decomposed as
\begin{align}
	\mathcal{H}_q = \mathcal{M}_q \otimes \mathcal{N}_q,
\end{align}
where $\mathcal{M}_q$ carries an irreducible representation (irrep) of $G$ and $\mathcal{N}_q$ is a multiplicity space, carrying a trivial representation of $G$. 
 
Thus, the Hilbert space can be decomposed as
\begin{align}
	\mathcal{H}=\bigoplus_q \mathcal{M}_q \otimes \mathcal{N}_q,
\end{align}
and, as proven in Ref.~\cite[II.C]{Bartlett_2007},
 a $(G,\mathcal{U})$-invariant operator $\tilde{O}$
 must be of the form
\begin{align}
\label{eq_opinv}
	\tilde{O}=\bigoplus_q c_q( \mathbb{1}_{\mathcal{M}_q} \otimes O_{\mathcal{N}_q}) ,
\end{align}
where $c_q\in \mathbb{R}$, $\mathbb{1}_{\mathcal{M}_q}$ is the identity operator on $\mathcal{M}_q$ and $O_{\mathcal{N}_q}$ is an arbitrary Hermitian operator on $\mathcal{N}_q$. Therefore, any $(G,\mathcal{U})$-invariant operator  must be block-diagonal relative to the $\mathcal{H}_q$ sectors (hence incoherent across the sectors),  and the reduction within each subsystem $\mathcal{M}_q$ must be maximally mixed. (The lack of coherence across sectors is what the term ``superselection'' traditionally refers to.)

In \cref{app:proofqt}, We illustrate with some examples the decomposition of the Hilbert space and the explicit form of invariant operators.  

It is not hard to show that the set of invariant states is tomographically complete for the set of invariant effects, and vice-versa. This will be seen as a special case of the argument in Appendix~\ref{app:tomographiccompleteness}.

\section{Proof of the failure of tomographic locality based on parameter counting}
\label{app:proofqt}

In the main text, all the proofs of the failure of tomographic locality have proceeded by finding explicit examples of sets of states that cannot be discriminated by local measurements. 
An alternative approach is to use the sufficient condition for the failure of tomographic locality given in Ref.~\cite{Hardy2011Limited}. Let $K_{A}$ denote the number of independent real parameters in an unnormalized state for a single system $A$  (equivalently, $K_A$ is the minimal cardinality of a tomographically complete set of effects), and let $K_{B}$ be defined analogously. 
Let $K_{AB}$ denote the number of independent real parameters in an unnormalized state 
for the composite system consisting of $A$ and $B$.

Then, the sufficient condition for the failure of tomographic locality for a GPT is 
\begin{align}
	\label{eq_hardy}
K_{A B} > K_{A} K_{B}.
\end{align}

In the following, we demonstrate how to use \cref{eq_hardy} explicitly in the rotation-twirled spinor world. This also provides us an opportunity to illustrate some of the 
concepts introduced in \cref{app_SSR}.

A single spinor 
carries an irrep of SU(2) with a trivial multiplicity space. As such, any rotationally invariant operator of a single spinor 
must be the identity operator up to a constant. Such an operator is specified by just one real parameter. 
For each of a pair of spinors, denoted $A$ and $B$, we consequently have
\begin{align}\label{KAKBrot}
	K_{A}= K_{B} = 1.
\end{align}

For the composite of a pair of spinors, $AB$, 
the Hilbert space decomposes as
\begin{align}
	\mathcal{H}_{AB} = \mathcal{H}_{1/2} \otimes \mathcal{H}_{1/2} = \mathcal{H}_{0} \oplus \mathcal{H}_{1},
\end{align}
where $\mathcal{H}_{j}$ denotes the Hilbert space of a spin-$j$. That is, the Hilbert space of the pair decomposes into the direct sum of a spin-0 subspace and a spin-1 subspace, each of which has a trivial multiplicity space. Therefore, any rotationally Hermitian invariant operator on a pair of spinors 
must be of the form
\begin{align}
	c_{0} \mathbb{1}_{0} + c_{1} \mathbb{1}_{1},
\end{align}
where $c_{0}, c_1\in\mathbb{R}$, $ \mathbb{1}_{0}$ is the identity operator on the spin-0 subspace, and $\mathbb{1}_{1}$ is the identity operator on the spin-1 subspace.  Clearly then, a rotationally invariant operator on $AB$ is specified by 2 real parameters,
\begin{align}\label{KABrot}
	K_{AB} =2.
\end{align}
Combining Eqs.~\ref{KAKBrot} and ~\ref{KABrot}, we conclude that
\begin{align}
	K_{A B}  > K_{A}K_{B},
\end{align}
and thus that the rotation-twirled spinor world exhibits a failure of tomographic locality.

\section{Introduction to generalized probabilistic theories}\label{definingGPTs}

We now briefly review the GPT framework~\cite{barrett2007information,Janotta_2014,Plavala_2023,muller2021probabilistic}. There are various slightly different ways to define GPTs; the version of the GPT framework presented below is adapted to our purposes. For simplicity, we limit our consideration to finite-dimensional tomographically local GPTs and their (not necessarily tomographically local) subtheories. 

A GPT \emph{system} $A$ is associated with a tuple $(R_A,\Omega_A,E_A, u_A,\mathcal{T}_A)$ of a real finite-dimensional vector space $R_A$, a convex set $\Omega_A$ of state vectors, a convex set $E_A$ of effect vectors with a chosen unit effect $u_A$, and a convex set $\mathcal{T}_A$ of transformations, with the following properties.

The \emph{state space} $\Omega_A$ is given by a convex set of vectors in $R_A$ whose affine span does not contain the origin. Note that, unlike many other approaches to formalizing GPTs, we will not assume that the linear span of $\Omega_A$ is $R_A$. This is in order to allow for the GPT to be a tomographically nonlocal subtheory of some tomographically local theory.  

Let $R_A^*$ be the space of linear functionals on $R_A$.  
For a single system, the \emph{effect space} $E_A$ is a  convex  subset of $R_A^*$  (termed {\em effects}) satisfying the following properties 
\begin{itemize} 
\item  every effect yields values between zero and one when acting on states, i.e., 
$ e(\omega) \in [0,1] $  for all  $e \in E_A$  and  \(\omega \in \Omega_A\);
\item  the set includes 
a chosen\footnote{Commonly, one considers GPTs wherein the states span $R_A$, in which case the unit effect is unique. Here we do not make this assumption, so there can be multiple candidates for a unit effect.  However, in all of our examples there is a canonical choice---namely, the unique unit effect in the original pre-twirled GPT. \blk} 
unit effect $u_A \in E_A$,  defined as a 
linear functional  such that  
\(u_A(\omega) = 1\) for all \(\omega \in \Omega_A\); 
\item  for every effect in the set, the complementary effect, defined as $e^\perp := u_A-e$, is also in the set, that is, $e \in E_A \implies e^\perp \in E_A$.\footnote{It follows that for every $e \in E_A$, there is a binary-outcome measurement, $\{e,e^\perp\}$, containing it and its complement.  Note that this condition alone is not sufficient to ensure that every set of effects which sums to the unit can be viewed as a valid measurement  (a property which is not typically demanded for a general GPT).  For example, it could be the case that $e_1+e_2+e_3+e_4=u$ but that $e_1+e_2$ is not in the effect space,  
and as it is always possible to coarse-grain outcomes of a measurement, this means that $\{e_1,e_2,e_3,e_4\}$ cannot be taken as a physically possible measurement in the theory. } 
\end{itemize}
Note the last two bullet points imply that the set includes the zero effect, which is 
a linear functional that assigns probability zero to all states and is denoted by $\mathbb{0}$, and that the second bullet point implies that $\Omega_A$ is the {\em normalized} state space. 
Because we wish to allow for failures of tomographic locality, we will not assume that the linear span of $E_A$ is $R^*_A$. 

As we are not making the common assumption that states and effects span $R_A$ and $R_A^*$ respectively, we must furthermore impose the following two conditions. 
For any two distinct states, $\omega, \omega'\in \Omega_A$ s.t. $\omega \neq \omega'$, there exists an effect $e\in E_A$ which distinguishes them, i.e., for which $e(\omega)\neq e(\omega')$.  Similarly, for any two distinct effects, $ e,e'\in E_A$ s.t. $e\neq e'$, there exists a state  $\omega \in \Omega_A$ which distinguishes them, i.e., for which $e(\omega)\neq e'(\omega)$. 
The states and effects are said to be {\em tomographic} (or {\em separating}) for each other, and hence GPTs are said to have the property of \emph{tomographic completeness}.

The \emph{transformation space} $\mathcal{T}_A$ is a convex set of linear maps on $R_A$ such that
\begin{itemize}
    \item normalized states are mapped to normalized states, i.e., \(T\circ \omega \in \Omega_A\) for all \(\omega \in \Omega_A\)    ;
    \item effects are mapped to effects (by precomposition), i.e., $e\circ T \in E_A$ for all $e\in E_A$;
    \item where, in particular, the unit effect is preserved, i.e., $u_A\circ T = u_A$.
    \item Moreover, the set of transformations is  closed under sequential composition, i.e., for all $T_1, T_2 \in \mathcal{T}_A$ we have that $T_1\circ T_2 \in \mathcal{T}_A$;
    \item and must contain the identity transformation, $\mathds{1}_A \in \mathcal{T}_A$, which is the unit for sequential composition.
\end{itemize}
Here $\circ$ simply represents sequential composition of linear maps. (If one chooses a particular basis for each vector space and obtains a matrix representation of these linear maps, then $\circ$ represents matrix multiplication.) Transformations can also map between different systems, but we will not need these in this article. Similarly, one often also considers transformations that are not normalization preserving but merely normalization non-increasing  (the analog of trace non-increasing transformations in quantum theory), but these will also not be needed here. 

A GPT can then be defined as a collection of GPT systems, which is closed under a suitably defined notion of composition.  For our purposes, it suffices to consider bipartite systems. Given a pair of systems $A=(R_A,\Omega_A,E_A, u_A,\mathcal{T}_A)$ and $B=(R_B,\Omega_B,E_B, u_B,\mathcal{T}_B)$ in the GPT, there must exist a composite system $AB=(R_{AB},\Omega_{AB}, E_{AB}, u_{AB}, \mathcal{T}_{AB})$ where (for the purposes of this paper) we take 
\begin{equation}\label{standardtensorproduct}
R_{AB}=R_A\otimes R_B.
\end{equation}
The composite state space must at least contain all separable states, i.e., $\mathsf{Conv}[\Omega_A\otimes\Omega_B]\subseteq\Omega_{AB}$, and similarly, the composite effect space must contain all separable effects $\mathsf{Conv}[E_A\otimes E_B]\subseteq E_{AB}$
\footnote{In fact, the composite effect space must also contain all effects which can be obtained from coarse-grainings of binary-outcome measurements performed on each system. For example, for a bipartite system if we have measurement $\{e,e^\perp\}$ on the first system, and measurement $\{f,f^\perp\}$ on the second, then we necessarily have the four outcome measurement $\{e\otimes f, e^\perp \otimes f, e\otimes f^\perp, e^\perp\otimes f^\perp\}$ and, hence, the effects $\alpha e\otimes f + \beta e^\perp \otimes f + \gamma e \otimes f^\perp + \delta e^\perp \otimes f^\perp$ must be in the bipartite effect space for arbitrary $\alpha,\beta,\gamma,\delta \in [0,1]$. 
Similarly, the composite effect space must also contain all effects which can be obtained from classical control of a measurement on one system by the outcome of the other. In the bipartite case this means, for example, that we have measurements such as $\{e\otimes f_1, e\otimes f_1^\perp, e^\perp \otimes f_2, e^\perp \otimes f_2^\perp\}$, where $f_1$ and $f_2$ are arbitrary effects on the second system, and hence all of the effects corresponding to linear sums of these with coefficients in $[0,1]$. Note that all of these extra effects can, however, be rescaled to convex combinations of product effects, and so do not provide any further ability to discriminate states over those in $\mathsf{Conv}[E_A\otimes E_B]$. } where $u_{AB}=u_A\otimes u_B\in E_{AB}$, and the composite transformation space must contain all separable transformations, $\mathsf{Conv}[\mathcal{T}_A\otimes\mathcal{T}_B]\subseteq \mathcal{T}_{AB}$.

We also have constraints coming from the composition of all of these elements. Of particular relevance for us are that ``steered'' states/effects are valid states/effects. More precisely, for all composite states $\omega^{AB}\in \Omega_{AB}$, it must be the case that when one composes this state with the effect $e^B$ on system $B$, the remaining system is in a valid state up to normalization. That is, noting that the set of normalized and subnormalized states are simply the convex hull of $\Omega_A$ and the zero state (which assigns probability zero to all effects), denoted $\mathtt{0}$, i.e.,
${\rm Conv}[\Omega_A \cup \{ \mathtt{0} \}]$, 
it must be the case that 
\beq
\label{eq:statesteer}(\mathds{1}^A\otimes e^B )\circ \omega^{AB} \in  
{\rm Conv}[\Omega_A \cup \{ \mathtt{0} \}]
\eeq
Similarly, for all composite effects $e^{AB}\in E_{AB}$, it must be the case that composition with a state $\omega_B$ of system $B$ leads to a valid effect on system $A$, i.e.,
\beq
\label{eq:effectsteer} e^{AB} \circ (\mathds{1}^A\otimes \omega^B) \in E_A.
\eeq 
We will refer to the pair of these constraints as `closure under steering'. 

There are further constraints on forming composite systems  which are often imposed but which we will not discuss here (see \cite{ying2024twirling2} for more details). In short, however, these extra conditions ensure that the GPT can be thought of as a symmetric monoidal category. In particular, in the formalism we introduce here, the GPT can be thought of as a subcategory of the category of real vector spaces and linear maps.

Here we provide a useful definition of tomographic locality which is equivalent to the definition given in our introduction, as shown in Ref.~\cite{Chiribella2010purification}.
\begin{definition}[Tomographic Locality]
A GPT is said to satisfy tomographic locality if and only if for any two distinct bipartite states $\omega_1^{A B}$ and $\omega_2^{A B}$, there exists a product effect $e^{A}\otimes e^{B}$ such that 
\begin{align}
\label{eq:LD}
(e^{A}\otimes e^{B})\circ \omega_1^{AB}   \neq  (e^{A}\otimes e^{B})\circ \omega_2^{A B}.   
\end{align}    
\end{definition}

Note that thanks to our assumption that the composite vector spaces are given by the tensor product (Eq.~\eqref{standardtensorproduct}), Eq.~\eqref{eq:LD} would automatically be satisfied if we additionally made the common assumption that $\mathsf{Span}(\Omega_{A})=R_{A}$ and $\mathsf{Span}(\Omega_{B})=R_{B}$.  Our relaxation of this condition allows for the kinds of tomographically nonlocal theories which are relevant for this project---namely, those which can be seen as subtheories of tomographically local ones. It remains an open question as to whether all tomographically nonlocal theories can be represented in this way.

\section{Symmetries and Twirling in arbitrary GPTs}
\label{app_formaldef}

It is straightforward to generalize the definition of a twirled world from quantum theory to an arbitrary GPT. 

We presume that as part of the specification of the GPT, there is a specification of the physical symmetries of the system, together with a specification of how these are represented on the GPT vector space $R_S$. Let $S=S_1 \dots S_n $ be a composite system, as before. 
Consider a physical symmetry associated to the group $G$ and let $\mathcal{V}^{S_i}$ denote the representation of this symmetry on system $S_i$, that is, $\mathcal{V}^{S_i}:G \to \mathcal{T}^{S_i}$.
Introducing the shorthand notation $\mathcal{V}^{S_i}_g:= \mathcal{V}^{S_i}(g)$, being a representation means that $\mathcal{V}^{S_i}_g\circ\mathcal{V}^{S_i}_{g'} = \mathcal{V}^{S_i}_{g\cdot g'}$ and $(\mathcal{V}^{S_i}_g)^{-1}=\mathcal{V}^{S_i}_{g^{-1}}$.  
The collective representation of this symmetry on the composite system $S$ is then  
$\mathcal{V}^{S}: G \to \mathcal{T}^{S_1} \otimes \mathcal{T}^{S_2} \dots \otimes \mathcal{T}^{S_n}$, 
where 
\begin{align}
\mathcal{V}^{S}_g := \mathcal{V}^{S_1}_g \otimes \dots \otimes \mathcal{V}^{S_n}_g.
\label{compositionGPT}
\end{align}
Note that if one formulates quantum theory as a GPT,  the GPT vector space is the real vector space of Hermitian operators (not the complex Hilbert space), so that the $\mathcal{V}^{S}$  are exactly the superoperator representations used in the main text. 

We say that a GPT state on system $S$, $\omega^{S}\in \Omega_{S}$ is {\em $(G,\mathcal{V})$-invariant} if and only if
\begin{align}
  \mathcal{V}^{S}_g \circ \omega^{S}= \omega^{S},\ \ \forall g\in G,
\end{align}
while a GPT effect on system $S$, $e^{S}\in E_S$, is said to be {\em $(G,\mathcal{V})$-invariant} if and only if
\begin{align}
  e^{S}\circ \mathcal{V}^{S}_g= e^{S},\ \ \forall g\in G.
\end{align}

A GPT operation from $S$ to itself, represented as a linear map on the GPT vector space $R_S$, denoted $T^{S} \in \mathcal{T}_S$, 
is said to be {\em $(G,\mathcal{V})$-covariant} if and only if
\begin{align}
 T^{S} \circ \mathcal{V}^{S}_g =  \mathcal{V}^{S}_g \circ  T^{S},\ \ \forall g\in G.
\end{align}
The generalization to the case where the operation goes from $S$ to some $S'$ is straightforward, but we will not need it here.

Then, a {\em $(G,\mathcal{V})$-twirled world} for a set of GPT systems is a subtheory of that GPT on those systems where the states and the effects are all and only those that are $(G,\mathcal{V})$-invariant and where the operations are all and only those that are $(G,\mathcal{V})$-covariant.

\subsection{The twirling map}

Again, it is straightforward to define a transformation that takes the set of all GPT states (by composition) and effects (by precomposition) to the subsets of these that are $(G,\mathcal{V})$-invariant.  Paralleling the terminology for the quantum case, we term this the {\em $(G,\mathcal{V})$-twirling} map. 
 It is defined, as before, by 
 group averaging:
\begin{align}
\mathcal{G}  := \int_G {\rm d}g\ \mathcal{V}^S_g, 
\label{twirlingmap}
\end{align}
where ${\rm d}g$ is the normalized Haar measure on $G$.

The GPT twirling map of Eq.~\eqref{twirlingmap} shares many of the same properties as the quantum twirling map.

\begin{lemma}\label{lem:TwirlingInvaraince}
     The GPT $(G,\mathcal{V})$-twirling map is invariant under right or left composition with any element of the representation  $\mathcal{V}$ of group $G$, that is,
\beq
\mathcal{G}\circ \mathcal{V}_g  = \mathcal{G} = \mathcal{V}_g \circ \mathcal{G} \ \ \forall g\in G.
\eeq
\end{lemma}
\begin{proof}
    This follows immediately from the fact that $\mathcal{V}$ is a representation and the invariance of the Haar measure:
    \beq
    \mathcal{G}\circ \mathcal{V}_g = \int_G {\rm d}g'\ \mathcal{V}_{g'} \circ \mathcal{V}_g = \int_G {\rm d}g' \mathcal{V}_{g'\cdot g} = \int_G {\rm d}g'' \mathcal{V}_{g''} = \mathcal{G}.
    \eeq
    where we have used the change of variable $g'':=g'\cdot g$ and the invariance of the Haar measure tells us that ${\rm d} g'' = {\rm d} g'$. The right invariance of the twirling map follows similarly.
\end{proof}

\begin{lemma}\label{glob2}  The GPT $(G,\mathcal{V})$-twirling map  is idempotent, that is
\beq 
\mathcal{G} \circ \mathcal{G} = \mathcal{G}.
\eeq
\end{lemma}
\begin{proof}
This follows immediately from Lemma~\ref{lem:TwirlingInvaraince}, and normalisation of the Haar measure
\beq
\mathcal{G} \circ \mathcal{G} = \int_G {\rm d}g\ \mathcal{V}_g \circ \mathcal{G} = \int_G {\rm d}g\ \mathcal{G} = \mathcal{G}.
\eeq
\end{proof}

\begin{lemma} \label{globloc}If we have a composite system $S=S_1\cdots S_n$ then  applying global twirling  on $S$ followed (or proceeded) by applying local twirling on  some subset of subsystems, is the same as applying local twirling on both this subset and the complementary subset. In the bipartite case (which is all that we need here) this means that:
\begin{align}
\mathcal{G}^{S_1}\otimes \mathcal{G}^{S_2} &= 
(\mathds{1}\otimes \mathcal{G}^{S_2})\circ \mathcal{G}^S\\
&= \mathcal{G}^S\circ (\mathds{1}\otimes \mathcal{G}^{S_2})\\
&= (\mathcal{G}^{S_1}\otimes \mathds{1})\circ \mathcal{G}^S\\
&= \mathcal{G}^S\circ (\mathcal{G}^{S_1}\otimes\mathds{1} )\\
&=  (\mathcal{G}^{S_1}\otimes \mathcal{G}^{S_2} )\circ \mathcal{G}^S\label{compo5}\\
&=  \mathcal{G}^S\circ (\mathcal{G}^{S_1}\otimes \mathcal{G}^{S_2} ).
\end{align} 
\end{lemma}
\begin{proof}
    For simplicity we just prove one of these equations, and the others follow similarly. It suffices to appeal to  Lemma~\ref{lem:TwirlingInvaraince} and the fact that $G$ acts collectively on the composite system $S$:
    \begin{align}
         (\mathcal{G}^{S_1} \otimes \mathds{1}) \circ \mathcal{G}^S &= (\mathcal{G}^{S_1} \otimes \mathds{1}) \circ  \int_{G}{\rm d}g\ \mathcal{V}^{S_1}_g \otimes \mathcal{V}^{S_2}_g \\
         &= \int_{G}{\rm d}g\ (\mathcal{G}^{S_1} \otimes \mathds{1}) \circ   (\mathcal{V}^{S_1}_g \otimes \mathcal{V}^{S_2}_g ) \\
         &= \int_{G}{\rm d}g\ (\mathcal{G}^{S_1}\circ \mathcal{V}^{S_1}_g) \otimes \mathcal{V}^{S_2}_g \\
         &= \int_{G}{\rm d}g\ \mathcal{G}^{S_1} \otimes \mathcal{V}^{S_2}_g  \\
         &= \int_{G}{\rm d}g\ (\mathcal{G}^{S_1} \otimes \mathbb{1}) \circ (\mathbb{1}\otimes\mathcal{V}^{S_2}_g)\\
        &= (\mathcal{G}^{S_1} \otimes \mathbb{1}) \circ 
        \int_{G}{\rm d}g\ (\mathbb{1}\otimes\mathcal{V}^{S_2}_g)\\ 
  &= \mathcal{G}^{S_1} \otimes \mathcal{G}^{S_2}.
    \end{align}
\end{proof}
\blk

\subsection{ Proof that a twirled world is a valid GPT \blk}

Throughout this work, we are viewing the twirled worlds that we construct as foil theories (GPTs) in their own right, as opposed to merely fragments~\cite{selby2023accessible} of the original theory (prior to twirling). 

For this to be a sensible view, it is important that the twirled worlds satisfy all of the properties of a GPT. Here we will check the basic conditions that the states and effects of a bipartite twirled world must satisfy, leaving for Ref.~\cite{ying2024twirling2} the full proof for a fully compositional twirled world (including, e.g., transformations). 

Most of the required properties are directly inherited from the original GPT.
For instance, the constraint that a state or effect be invariant under the twirling map is linear, and so it follows immediately that the twirled world's state and effect spaces are convexly closed and that the composition of states and effects gives valid probabilities.\footnote{In fact, all that is needed for these to follow is that the twirling map is \emph{real}-linear instead of complex-linear.} Moreover, the fact that the twirling operation preserves the unit effect implies that for each effect in the twirled theory, the complementary effect is also in the twirled theory (indeed, one can immediately check that 
the complement of the image under the twirling map of an effect is the image under the twirling map of the complement of that effect.

The two slightly less trivial conditions which must be checked are that: (i) twirled states and twirled effects are tomographic for one another, and
(ii) that the state (effect) space of multipartite systems is closed under composition with effects (states) on subsystems, i.e., that the closure under steering conditions \cref{eq:statesteer} and \cref{eq:effectsteer} are satisfied. 

\subsubsection{Proof that a twirled world is tomographically complete}
\label{app:tomographiccompleteness}

We now prove that the states and effects in a twirled world are tomographic for each other, and that this applies also to composite systems, so that bipartite states that cannot be distinguished by local measurements (due to tomographic nonlocality) are still distinguishable by some global measurements on the composite.

Consider a GPT and its twirling relative to some $(G,\mathcal{V})$-twirling map. We wish to show that any two distinct invariant states, $\omega_1 \neq \omega_2$ can be distinguished by some invariant effect.

Since $\omega_1$ and $\omega_2$ are distinct, there exists (by definition) an effect $e$ in the original, pre-twirled
GPT such that
\begin{align}
\label{neqev}
    e \circ \omega_1 \neq e \circ \omega_2.
\end{align}

Since $\omega_1$ and $\omega_2$ are invariant, by definition, they satisfy that 
\begin{align}
\label{inv12}
    \omega_1 = \mathcal{G} \circ  \omega_1, \quad \omega_2 = \mathcal{G} \circ \omega_2.
\end{align}

Combining \cref{neqev} and \cref{inv12}, we have
\begin{align}
\label{neqeG}
    e \circ \mathcal{G} \circ \omega_1 \neq e \circ \mathcal{G} \circ \omega_2.
\end{align}

It follows that 
\begin{align}
    e \circ \mathcal{G}
\end{align}
is an effect that separates $\omega_1$ and $\omega_2$. And $ e \circ \mathcal{G}$ is evidently invariant under  $\mathcal{G}$.

Thus, for any two distinct states $\omega_1$ and $\omega_2$ in the twirled world arising from imposing $(G,\mathcal{V})$-symmetry on the GPT, there exists an effect, namely, $ e \circ \mathcal{G} $ in that twirled world that distinguishes the two.

A similar argument can be used to prove that in such a twirled world, for any two distinct effects, there exists a state that can distinguish the two.

\subsubsection{Proof that a twirled world is closed under steering}

As in the discussion surrounding Eqs.~\ref{eq:statesteer}-\ref{eq:effectsteer}, it must be the case for any valid GPT that when we have a bipartite state (resp. effect) and compose it with a local effect (resp. state), we obtain a valid local state (resp. effect).

To see that Eq.~\ref{eq:statesteer} is satisfied in any twirled world, consider a bipartite state $\omega^{AB}$ and an effect $e^B$ on system $B$, both of which live in a given twirled world. When we compose these we obtain $\omega^A :=(\mathds{1}^A\otimes e^B) \circ \omega^{AB}$,  and we need to show that this is a valid state in the twirled world---i.e., that it satisfies $\omega^A = \mathcal{G}^A \circ \omega^A$. This follows from Lemma~\ref{globloc} and the fact that $\omega_{AB}$ and $e^B$ belong to the twirled world, so that $\omega^{AB} = \mathcal{G}^{AB}\circ \omega^{AB} $  and $e^B = e^B \circ \mathcal{G}^B$. It then follows that
\begin{align}
\omega^A&:=(\mathds{1}^A\otimes e^B) \circ \omega^{AB}\\
&= (\mathds{1}^A\otimes e^B) \circ (\mathds{1}^A \otimes \mathcal{G}^B) \circ \mathcal{G}^{AB} \circ \omega^{AB} \\
&\stackrel{\text{Lem.}~\ref{globloc}}{=}(\mathds{1}^A \otimes e^B) \circ (\mathcal{G}^A\otimes \mathcal{G}^B) \circ \omega^{AB} \\
&= (\mathds{1}^A \otimes e^B) \circ (\mathcal{G}^A \otimes \mathds{1}^B) \circ \omega^{AB} \\
&= \mathcal{G}^A \circ \omega^A.
\end{align}
Condition \ref{eq:effectsteer} has a similar proof.

\section{Proof of the ubiquity theorem}\label{app:ubiquity}

To prove \cref{thm:ubiquity}, we first need to use a specific reparameterization. For any system \( S \), it is possible to reparametrize the states and effects using any linear invertible transformation \( L: R_S \to R_S \), where states transform as \( \omega^S \to L \circ \omega^S \) and effects as \( e^S \to e^S \circ L^{-1} \), ensuring that probabilities remain unchanged:  
\[
\left( e^S \circ L^{-1} \right) \left(L \circ \omega^S \right) = e^S (\omega^S).
\]  
Similarly, for transformations, one has \( T^S \to L \circ T^S \circ L^{-1} \), preserving the dynamical structure of the system:  
\[
\left(L \circ T^S \circ L^{-1} \right) \circ \left(L \circ \omega^S \right) = L \circ \left( T^S \circ \omega^S \right).
\]
It is a standard fact about the 
representation of compact groups that they 
can be reparametrized so as to become unitary; see, e.g., \cite{Vinberg}. For completeness, we include a short proof, 
due to~\cite{Masanes_2013}.

\begin{lemma}
\label{Orthogonality}
For any compact Lie group \( G \), including finite groups, 
one can reparametrize any representation $\mathcal{V}^S$ to an orthogonal representation $\tilde{\mathcal{V}}^S$, 
\begin{equation}
\left({\tilde{\mathcal{V}}^{S}_{g}}\right)^T \circ\tilde{\mathcal{V}}^{S}_{g} =\mathds{1}.
\end{equation}
\end{lemma}
\begin{proof}

Since $G$ is a compact Lie group there exists a unique normalized (left- and right-invariant) Haar measure $dg$~\cite{bump_lie_2013}. We define

\begin{equation}
    W^2: = \int_{g \in G} dg \  \left(\mathcal{V}^{S}_{g}\right)^{T}\circ \mathcal{V}^{S}_{g}.
\end{equation}
$\mathcal{V}^{S}_{g}: R_S \to R_S$ is invertible hence $\mathcal{V}^{S}_{g} v \neq 0$ for any $v \in R_S$, $v \neq 0$. Denote the orbit $v(g) = \mathcal{V}^{S}_{g} v$, where $v \in R_S$, for all $g \in G$. Then $v^T W^2 v = \int_G dg v(g)^T v(g) = \int_G \langle v(g) , v(g) \rangle$. For $v \neq 0$, $v(g) \neq 0$ by the above and hence $\langle v(g) , v(g) \rangle >0$. This implies that $W^2$ is a positive definite operator. 
We then define \( W \) as its unique positive square root. Additionally, note that \( W^T = W \).

Now, using \( W \) as the linear map \( L = W \) to reparametrize the states and effects, and subsequently reparametrizing \( \mathcal{V}^{S}_{g} \) as \( \tilde{\mathcal{V}}^{S}_{g} \), we obtain  
\begin{align}
    &\left({\tilde{\mathcal{V}}^{S}_{g}}\right)^T\circ \tilde{\mathcal{V}}^{S}_{g} \nonumber\\
    &= \left( W \circ \mathcal{V}^{S}_{g} \circ W^{-1} \right)^T \circ
       \left( W \circ \mathcal{V}^{S}_{g} \circ W^{-1} \right) \nonumber\\
    &=  W^{-1} \circ \left(\mathcal{V}^{S}_{g} \right)^T\circ\left(\int_{G} dg' \left(\mathcal{V}^{S}_{g'}\right)^{T}\circ 
       \mathcal{V}^{S}_{g'}\right) \circ\mathcal{V}^{S}_{g}\circ W^{-1}\nonumber \\
    &= W^{-1} \circ W^2\circ  W^{-1} = \mathds{1}.
\end{align}
\end{proof}
When clear from context, and without loss of generality, we use $\mathcal{V}^{S}$ to denote the reparametrized orthogonal representation as well.

Next, using the orthogonal reparametrized representation above, we show that global and local twirling are distinct linear maps. That is,
\begin{lemma}
For any composite system $AB$, with identical subsystems $A$ and $B$, each carrying the same non-trivial representations $\mathcal{V}^A$ and $\mathcal{V}^B$, we have that $\mathcal{G}^A\otimes \mathcal{G}^B \neq \mathcal{G}^{AB}$.
\end{lemma}
\proof
To prove this we show that the vector $c:=\sum_i\phi^{A}_i\otimes\phi^{B}_i$ for orthonormal basis $\phi_i^{A}$ and $\phi_i^{B}$ is invariant under $\mathcal{G}^{AB}$ but not under $\mathcal{G}^A\otimes\mathcal{G}^B$. The key to this is the fact that one has 
\begin{equation}
    (\mathds{1}_B\otimes \mathcal{V}^B_g)(c)=((\mathcal{V}^A_g)^T\otimes\mathds{1}_B)(c)=(\mathcal{V}^A_{g^{-1}}\otimes\mathds{1}_B)(c) .
\end{equation}
 where the first equality is easily verified by writing all of the linear maps as matrices in the $\phi_i$ orthonormal basis\footnote{To be more explicit, using braket-like notation for operators we can prove this as $\sum_i \mathds{1}\otimes L |\phi_i\phi_i) = \sum_{ijk}|\phi_i) L_{jk}|\phi_j)(\phi_k|\phi_i)= \sum_{ij}L_{ji}|\phi_i\phi_j)=\sum_{ij}(L^T)_{ij}|\phi_i\phi_j)=\sum_{ij}(L^T)_{ji}|\phi_j\phi_i)=\sum_{ijk}(L^{T})_{jk}|\phi_j)(\phi_k|\phi_i)|\phi_i)=\sum_i L^T\otimes \mathds{1}|\phi_i\phi_i)$.}, and the second is immediate from the
orthogonality of $\mathcal{V}^A_g$ and $\mathcal{V}^B_g$.  

Invariance under $\mathcal{G}^{AB}$ then immediately follows as
\begin{align}
\mathcal{G}^{AB}(c)&=\int_{g\in G}dg (\mathcal{V}_g^A\otimes\mathcal{V}_g^B)(c)\nonumber\\
&= \int_{g\in G}dg ((\mathcal{V}_g^A\circ \mathcal{V}_{g^{-1}}^{{A}})\otimes \mathds{1}_{B})(c)\nonumber\\
&=\int_{g\in G}dg\ c = c
\end{align}
where in the last step we are using normalisation of the Haar measure.

Next consider locally twirling one side of the vector $c$
\begin{align}\mathds{1}_A\otimes \mathcal{G}^B(c)& = \int_{g\in G}dg(\mathds{1}_A\otimes \mathcal{V}^B_g)(c)\nonumber\\&=\int_{g\in G}dg(\mathcal{V}^A_{g^{-1}}\otimes\mathds{1}_B)(c)\nonumber\\
&=\int_{g\in G}dg(\mathcal{V}^A_{g}\otimes\mathds{1}_B)(c)\nonumber=\mathcal{G}^A\otimes \mathds{1}_B(c).
\end{align}
Where we used the fact that the Haar measure is inversion invariant for compact Lie groups~\cite{bump_lie_2013}. Then, as twirling is idempotent we immediately find that
\[\mathcal{G}^A\otimes\mathcal{G}^B(c)=(\mathcal{G}^A)^2\otimes\mathds{1}_B(c)=\mathcal{G}^A\otimes\mathds{1}_B(c).\]

As the $\phi_i^A$ and $\phi_i^B$ form an orthonormal basis, we have that 
\[L\otimes \mathds{1}_B(c)=c \quad \iff \quad L=\mathds{1}_A\]
hence, $c$ is invariant under local twirling iff $\mathcal{G}^A=\mathds{1}_A$. However, by assumption, the representation is nontrivial such that $\mathcal{G}^A\neq \mathds{1}_A$, and so $c$ is not invariant under local twirling, just as we needed to show.
\endproof

Since $\mathcal{G}^A\otimes \mathcal{G}^B \neq \mathcal{G}^{AB}$ it follows that for any set of spanning vectors $\{v_i\}_i$ there must be at least one $v_i$ such that $(\mathcal{G}^A\otimes \mathcal{G}^B)(v_i) \neq \mathcal{G}^{AB}(v_i)$. Since the states span the underlying vector space, the vectors $v_i$ can always be chosen to be states, hence there is always at least one state $s$ such that
\begin{equation}
\label{sglobloc}
s_{\text{loc}}:=\mathcal{G}^A\otimes\mathcal{G}^B(s)\neq \mathcal{G}^{AB}(s):=s_{\text{glob}} .
\end{equation}
The existence of the two states, $s_{\text{loc}}$ and $s_{\text{glob}}$, ultimately leads to the following:
\begin{restatable}{theorem}{ubiquity}\label{thm:ubiquity}
For any GPT $\mathds{T}$ (not necessarily classical or quantum) and for any physical symmetry group $G$ with an action $\mathcal{V}$ that is nontrivial on some state(s)
within $\mathds{T}$, the corresponding $(G,\mathcal{V})$-twirled $\mathds{T}$-world exhibits a failure of tomographic locality.
\end{restatable}
\begin{proof}
First, we note that both states $s_{\text{glob}}$ and $s_{\text{loc}}$ in \eqref{sglobloc}, are $(G,\mathcal{V})$-invariant, i.e.,
\begin{align}
\label{eq:invS}
   s_{\text{loc}}= \mathcal{G}^{AB} \circ  s_{\text{loc}},\\
    s_{\text{glob}}= \mathcal{G}^{AB} \circ s_{\text{glob}},
\end{align}
a fact that follows in a straightforward manner from their definitions and Lemmas \ref{glob2} and \ref{globloc}. Hence, both are states within the twirled world. Moreover, because they are not equal and because the twirled world is tomographically complete (Appendix~\ref{app:tomographiccompleteness}), we know that they are {\em distinguishable} states within the twirled world.

In the remainder of the proof, we will show that these two states cannot be distinguished by local measurements, establishing that the twirled world is tomographically nonlocal.

 To assess tomographic locality, one must consider $(G,\mathcal{V})$-invariant effects of the product form.
 Let $e_1^{A}$ and $e_2^{B}$ be  an arbitrary pair of $(G,\mathcal{V})$-invariant effects,  
i.e.,  $ e_1^{A}\circ \mathcal{G}^A= e_1^{A}$ and $ e_2^{B}\circ \mathcal{G}^B= e_2^{B}$.   
The product effect these define is clearly invariant under local twirling 
\begin{align}
\label{eq:eeinv}
    (e_1^{A} \otimes e_2^{B} ) \circ (\mathcal{G}^A\otimes \mathcal{G}^B)= e_1^{A} \otimes e_2^{B}.
\end{align}
This product effect is also invariant under global twirling
\begin{align}
\label{eq:eeinv2}
    (e_1^{A} \otimes e_2^{B} ) \circ \mathcal{G}^{AB}= e_1^{A} \otimes e_2^{B}.
\end{align}
The proof is as follows
\begin{align}
    &(e_1^{A} \otimes e_2^{B} ) \circ \mathcal{G}^{AB} \\
    &=(e_1^{A} \otimes e_2^{B} ) \circ (\mathcal{G}^A\otimes \mathcal{G}^B) \circ \mathcal{G}^{AB} \\
    &=(e_1^{A} \otimes e_2^{B} ) \circ (\mathcal{G}^A\otimes\mathcal{G}^B)\\
    &=e_1^{A} \otimes e_2^{B},
\end{align}
 where  the first and last equalities follow from \cref{eq:eeinv} and the second equality follows from \cref{globloc} (specifically, \cref{compo5}). 

We are now in a position to show that no $(G,\mathcal{V})$-invariant product effect can distinguish $s_{\text{loc}}$ from $s_{\text{glob}}$, and hence that tomographic locality fails.  
It suffices to note that the probability assigned to any $(G,\mathcal{V})$-invariant product effect by the state $s_{\text{loc}}$ is the same as the probability assigned to it by the state $s$, and similarly for $s_{\text{glob}}$. More precisely, we have 
\begin{equation}\label{eq:jjj1}
    ( e_1^{A} \otimes e_2^{B} ) \circ s_{\text{loc}}
   =( e_1^{A} \otimes e_2^{B} )\circ s,
   \end{equation}
a result that follows directly from Eqs.~\eqref{sglobloc} and \eqref{eq:eeinv},   
   and we have
   \begin{equation}\label{eq:jjj2}
   ( e_1^{A} \otimes e_2^{B} ) \circ s_{\text{glob}}=( e_1^{A} \otimes e_2^{B} ) \circ s,
\end{equation}
a result that follows directly from Eqs.~\eqref{sglobloc} and \eqref{eq:eeinv2}.
Eqs.~\eqref{eq:jjj1} and \eqref{eq:jjj2} imply that 
\begin{equation}
    ( e_1^{A} \otimes e_2^{B} ) \circ s_{\text{loc}}= ( e_1^{A} \otimes e_2^{B} ) \circ s_{\text{glob}},
\end{equation}
that is, the probability assigned to any $(G,\mathcal{V})$-invariant product effect by the state $s_{\text{loc}}$ is the same as the probability assigned to it by $s_{\text{glob}}$.
We have therefore established that there is a pair of distinct  $(G,\mathcal{V})$-invariant states on $AB$ that cannot be discriminated by any $(G,\mathcal{V})$-invariant product effect, hence that there is a failure of tomographic locality.

\end{proof}

By requiring the underlying vector space $R_A$ of a system $A$ as being the span of its states $\omega_A$ one can concisely prove the result as follows. In this context for theories that do not satisfy tomographic locality, one has $R_{AB} \neq R_{A} \otimes R_{B}$. For the $(G,{\cal V})$-twirled world, $R_{AB}$ is spanned by the states $\omega_{AB}$ that are $(G,{\cal V})$-invariant whereas $R_A \otimes R_B$ is spanned by the states $\omega_A \otimes \omega_B$ which are invariant under $G \times G$. The state $s_{\rm glob}$ is $(G,{\cal V})$ invariant and hence in $R_{AB}$; however, it is not invariant under $G \times G$ and hence is not in $R_A \otimes R_B$. This implies that $R_{AB} \neq R_{A} \otimes R_{B}$. Thus, the $(G,{\cal V})$-twirled world violates tomographic locality.

There is an alternative (albeit less direct) way to demonstrate a failure of tomographic locality in a GPT. As noted in Ref.~\cite{chiribella2021process}, it suffices to demonstrate the existence of a pair of transformations on some system, say $A$, that are distinguishable when acting on states of a composite $AB$ but are indistinguishable when acting on states of $A$ alone. The proof provided above can be easily recast in this form. Specifically, consider the following pair of $(G,\mathcal{V})$-covariant transformations: the identity transformation $\mathds{1}^{A}$ and the twirling transformation $\mathcal{G}^{A}$. 

It is trivial to see that they both take any $(G,\mathcal{V})$-invariant state of $A$ to itself and therefore are indistinguishable when acting on system $A$ alone in the twirled world. However, they {\em can} be distinguished by their action on $(G,\mathcal{V})$-invariant states of the composite system $AB$.  It suffices to let $B$ be of the same type as $A$ and consider the $(G,\mathcal{V})$-invariant state $s_{\text{glob}}$ defined in Eq.~\eqref{sglobloc}, and then notice that
\beq
\mathcal{G}^A\otimes\mathds{1}^B(s_{\text{glob}}) =s_{\text{loc}},
\eeq
while
\beq
\mathds{1}^A\otimes \mathds{1}^B (s_{\text{glob}}) = s_{\text{glob}},
\eeq
and as noted earlier, $  s_{\text{loc}}\ne s_{\text{glob}}$.

\section{Post-quantum example: Reflection-twirled boxworld}\label{app:nonquantumexamples}

One of the most prominent examples of a post-quantum foil theory is boxworld~\cite{barrett2007information}, which was motivated by the desire to have a GPT that was non-signaling but allowed for maximal violations of Bell inequalities, that is, a GPT that incorporated the correlations known as PR boxes~\cite{Popescu1994PRbox}.\footnote{The terminology of ``twirled worlds'' introduced in this article is inspired by the use of ``boxworld'' as a pithy shorthand for the generalized no-signaling theory of Ref.~\cite{barrett2007information}.}
 Its construction allows one to have states that exhibit stronger forms of entanglement than are found in quantum theory, but at the cost of there being no entangled measurements.  
The lack of entangled measurements implies that boxworld trivially satisfies tomographic locality.
However, as we will show now, when we demand a reflection symmetry in boxworld, thereby defining the reflection-twirled boxworld, we induce a failure of tomographic locality.

We begin by providing a formal description of systems in boxworld. 

\begin{figure}[htbp]
 \begin{center}
 \includegraphics[width=0.5\textwidth]{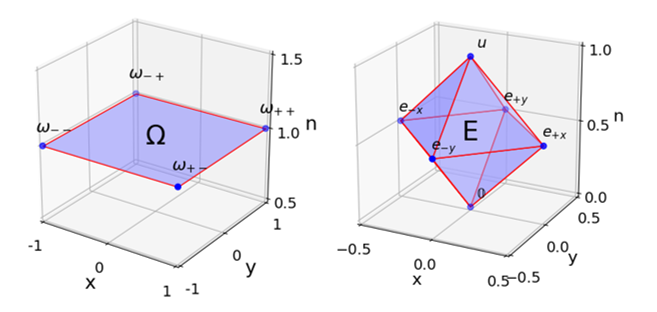}
 \caption{Left: the state space of a gbit. Right: the effect space of a gbit. Here, the two axes corresponding to nontrivial measurements are labelled ${x}$ and ${y}$, while the axis corresponding to normalization is labelled ${n}$. 
 }
 \label{gbit}
 \end{center}
 \end{figure}

The fundamental systems are termed generalized bits or ``gbits''\cite{barrett2007information}.  
The convex set of normalized states for a gbit is a square.
It is convenient to take the four convexly extremal (i.e., pure) states, which are the vertices of the square, to be the following four column vectors in $\mathds{R}^3$:
 \begin{align}
\omega_{++}&=(1,1,1)^T, \\
\omega_{+-}&=(1,-1,1)^T,\\
\omega_{-+}&=(-1,1,1)^T,\\  
 \omega_{--}&=(-1,-1,1)^T.
 \end{align}
The full state space of a gbit is the convex hull of these
\begin{equation}\label{boxworldstatespace}
	 \Omega = {\sf Conv} \left[\left\{ \omega_{++}, \omega_{+-}, \omega_{-+}, \omega_{--} \right\}\right],
 \end{equation}
or equivalently, 
$$\Omega =\left\{ (x,y,1)^T \in \mathds{R}^3 | -1\leq x\leq 1, -1\leq y \leq 1\right\}.$$ 
A depiction of this state space, embedded in $\mathbb{R}^3$, is given in \cref{gbit}.

Because boxworld is defined as satisfying the no-restriction hypothesis \cite{Chiribella2010purification}, every linear functional that yields valid probabilities on all states is included in the effect space, i.e., the theory includes every linear functional $e$ such that $e\circ \omega \in [0,1]$ for all $\omega \in \Omega$. 
The effect space for a gbit can therefore be deduced from the state space $\Omega$.

In the following, we will represent the linear functional $e$ as a row vector $e \in \mathbb{R}^3$ such that the action of $e$ on $\omega$, i.e., $e \circ \omega$, is simply the matrix product of the row vector $e$ with the column vector $\omega$. Any vector written inline is to be interpreted as a row vector, so that effect vectors can be written in-line as $(e_x,e_y,e_n)$, while state vectors are written in-line as $(\omega_x,\omega_y,\omega_n)^T$.

As usual, the effect space includes the unit effect, which is the linear functional $u$ such that $u \circ \omega:= 1$ for all $\omega \in \Omega$, which by virtue of \cref{boxworldstatespace} is represented by the row vector
$u \coloneq (0,0,1).$
It also includes the zero effect, denoted $\mathbb{0}$ and defined as the linear functional such that $\mathbb{0} \circ \omega:= 0$ for all $\omega \in \Omega$. Again, by virtue of \cref{boxworldstatespace}, this is represented by the row vector
$\mathbb{0} \coloneq (0,0,0).$
The full effect space in boxworld also includes the pair of linear functionals associated to row  vectors
 \begin{align}
  	  e_{+x} &\coloneq \tfrac{1}{2}(1,0,1), \\ 
	  e_{+y} &\coloneq \tfrac{1}{2}(0,1,1),
 \end{align}
and, as is required by the definition of an effect space, the complements of these
 \begin{align}
   	e_{-x} &\coloneq u -e_{+x} = \tfrac{1}{2}(-1,0,1), \\ 
	  e_{-y} &\coloneq u - e_{+y} = \tfrac{1}{2}(0,-1,1).
 \end{align}

The full effect space of a gbit is the convex hull of these six linear functionals
 \begin{equation}
 E \coloneq {\sf Conv}\left[\left\{ u, \mathbb{0},e_{+x}, e_{-x}, e_{+y}, e_{-y} \right\}\right].
 \end{equation}
It is also depicted in \cref{gbit}.

We now describe the state and effects spaces for a {\em pair} of gbits.  

The state spaces in boxworld compose using the \emph{max-tensor product}, that is, the state space of a composite includes all vectors in the tensor product vector space which give valid probabilities for all product effects. Concretely, for a pair of gbits, denoted $A$ and $B$, this state space is  the convex hull of the 16 product states and the eight {\em PR-box states}, which are defined as follows:
 \begin{align}
		 \omega^{AB}_{1}&=(1,1,0,1,-1, 0,0,0,1)^T,\\
		  \omega^{AB}_{2}&=(1,1,0,-1,1, 0,0,0,1)^T,\\
		  \omega^{AB}_{3}&=(1,-1,0,1,1, 0,0,0,1)^T,\\
		  \omega^{AB}_{4}&=(-1,1,0,1,1, 0,0,0,1)^T,\\
		  \omega^{AB}_{5}&=(-1,-1,0,-1,1, 0,0,0,1)^T,\\
		  \omega^{AB}_{6}&=(-1,-1,0,1,-1, 0,0,0,1)^T,\\
		  \omega^{AB}_{7}&=(-1,1,0,-1,-1, 0,0,0,1)^T,\\
		  \omega^{AB}_{8}&=(1,-1,0,-1,-1, 0,0,0,1)^T.
	 \label{PRs}
 \end{align}
 That is, 
 \begin{align}
\Omega_{AB} \coloneq {\sf Conv}\left[ \left\{ \omega^A_i \otimes \omega^B_j : i,j\in \{++,+-,-+,--\}\right\} \nonumber  \right. \\ \left. 
\cup \left\{ \omega^{AB}_{k}: k\in\{1,\dots,8\} \right\} \right].
 \end{align}

Because boxworld satisfies the no-restriction hypothesis,
every linear functional which yields valid probabilities on all states of a composite system is included as an effect on that composite system. It can be shown that this means that the effect space has no entangled effects, that is, that effects are necessarily constrained to certain positive linear combinations of product effects.

With the formal description of boxworld in hand, we can turn to defining the theory one obtains from boxworld by imposing a reflection-twirling. 

The symmetry group is $\mathbb{Z}_2$ where the pair of elements are the identity transformation, denoted $\mathbb{1}$, and the reflection transformation, denoted $r$. We consider the case of a reflection about the plane defined by the $y$ axis and the axis corresponding to normalization (denoted ${n}$ in Fig.~\ref{gbit}) in the vector space of a single gbit.  The action on a state $\omega$ is consequently a flipping of the sign of the $x$-component of $\omega$. Thus, the relevant representation of the symmetry is 
 \begin{equation}
\setstacktabbedgap{2pt}
\mathcal{V}_{\mathbb{1}}  = \parenMatrixstack{
1 & 0 & 0\cr
0 & 1 & 0\cr
0 & 0 & 1}
 \quad
 \mathcal{V}_r = \parenMatrixstack{
-1 & 0 & 0\cr
0 & 1 & 0\cr
0 & 0 & 1}.
 \end{equation}

\begin{figure}[htbp]
 \begin{center}
 \includegraphics[width=0.5\textwidth]{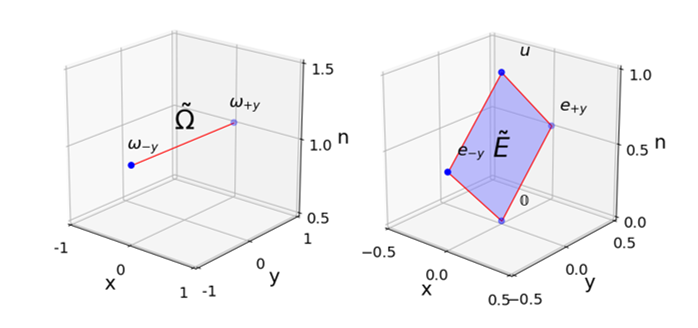}
 \caption{Left: the state space of a reflection-twirled gbit. Right: the effect space of a reflection-twirled gbit. }
 \label{twirledspaces}
 \end{center}
 \end{figure}

For a single gbit, the states that are invariant under this symmetry are those for which the component along the ${x}$-axis is zero.  This is simply  
the slice of the full state space (depicted in Fig.~\ref{gbit}) in the plane defined by the ${y}$-axis and the normalization axis, which corresponds to the line segment depicted in the left plot of Fig.~\ref{twirledspaces}. Clearly, there are two convexly extremal (i.e., pure) invariant states, namely,
   \begin{align}
   \omega_{+y} &= \tfrac{1}{2}\omega_{++} + \tfrac{1}{2}\omega_{-+}= (0,1,1)^T,\\
   \omega_{-y} &= \tfrac{1}{2}\omega_{+-} + \tfrac{1}{2}\omega_{--} = (0,-1,1)^T.
 \end{align}
Then, the full set of invariant states for a single gbit is the convex hull of this pair, 
\begin{equation}
\tilde{\Omega} = {\sf Conv}[\{ \omega_{+y}, \omega_{-y} \}].
\end{equation}

It is a similar story for the effects.  Those that are invariant under the reflection symmetry are those for which the corresponding vector has $x$-component zero. This is again simply the slice of the full effect space (depicted in Fig.~\ref{gbit}) in the plane defined by the ${y}$-axis and the normalization axis, which corresponds to the diamond shape depicted in the right plot of Fig.~\ref{twirledspaces}. The convexly extremal effects having $x$-component zero are the unit effect $u$, the zero effect $\mathbb{0}$, and the effects $e_{+y}$ and $e_{-y}$. 
Then, the full set of invariant effects for a single gbit is just the convex hull of these, 
\begin{equation}
\tilde{E} = {\sf Conv}[\{ u, 0, e_{+y}, e_{-y} \}].
\end{equation}

We now turn to the invariant states and effects for a pair of gbits, $A$ and $B$.  These are the states and effects that are invariant under the collective action of the group. Following Eq.~\eqref{compositionGPT} the collective representation on $AB$ is 
\begin{equation}
\setstacktabbedgap{2pt}
\begin{array}{cc}
&\mathcal{V}^{AB}_{\mathbb{1}}  = \mathcal{V}^{A}_{\mathbb{1}} \otimes \mathcal{V}^{B}_{\mathbb{1}} 
= \parenMatrixstack{
 1 & 0 & 0\cr
 0 & 1 & 0\cr
 0 & 0 & 1} 
 \otimes  \parenMatrixstack{
 1 & 0 & 0\cr
 0 & 1 & 0\cr
 0 & 0 & 1}\\
 
     & = \parenMatrixstack{
 1 & 0 & 0 & 0 & 0 & 0 & 0 & 0 & 0\cr
 0 & 1 & 0 & 0 & 0 & 0 & 0 & 0 & 0\cr
 0 & 0 & 1 & 0 & 0 & 0 & 0 & 0 & 0\cr
 0 & 0 & 0 & 1 & 0 & 0 & 0 & 0 & 0\cr
 0 & 0 & 0 & 0 & 1 & 0 & 0 & 0 & 0\cr
 0 & 0 & 0 & 0 & 0 & 1 & 0 & 0 & 0\cr
 0 & 0 & 0 & 0 & 0 & 0 & 1 & 0 & 0\cr
 0 & 0 & 0 & 0 & 0 & 0 & 0 & 1 & 0\cr
 0 & 0 & 0 & 0 & 0 & 0 & 0 & 0 & 1},\cr
\end{array}
\end{equation}

\begin{equation}
\setstacktabbedgap{2pt}
\begin{array}{cc}     
&\mathcal{V}^{AB}_r  = \mathcal{V}^{A}_r \otimes \mathcal{V}^{B}_r 
= \parenMatrixstack{
 -1 & 0 & 0\cr
 0 & 1 & 0\cr
 0 & 0 & 1} 
 \otimes
 \parenMatrixstack{
 -1 & 0 & 0\cr
 0 & 1 & 0\cr
 0 & 0 & 1}\\
 &= \parenMatrixstack{
 1 & 0 & 0 & 0 & 0 & 0 & 0 & 0 & 0\cr
 0 & -1 & 0 & 0 & 0 & 0 & 0 & 0 & 0\cr
 0 & 0 & -1 & 0 & 0 & 0 & 0 & 0 & 0\cr
 0 & 0 & 0 & -1 & 0 & 0 & 0 & 0 & 0\cr
 0 & 0 & 0 & 0 & 1 & 0 & 0 & 0 & 0\cr
 0 & 0 & 0 & 0 & 0 & 1 & 0 & 0 & 0\cr
 0 & 0 & 0 & 0 & 0 & 0 & -1 & 0 & 0\cr
 0 & 0 & 0 & 0 & 0 & 0 & 0 & 1 & 0\cr
 0 & 0 & 0 & 0 & 0 & 0 & 0 & 0 & 1}.
 \end{array}
\end{equation}
Note that these two matrices differ in sign in the second, third, fourth and seventh elements of the diagonal. Thus, the states on a pair of gbits that are invariant under reflection symmetry are all and only those with a 0 in the second, third, fourth and seventh components of their vector representation. 

 It follows that the bipartite effects on $AB$ which are invariant under the collective symmetry are all and only those in the convex hull of the following effects: the zero effect, the nine products of invariant nonzero effects, i.e., those in the set 
$\{e^A_{+y},e^A_{-y}, u^A\} \otimes \{e^B_{+y},e^B_{-y}, u^B\}$,
 plus two more invariant effects, namely,
  \begin{align}
  	  e^{AB}_+ &= \tfrac{1}{2}[e^A_{+x}\otimes e^B_{+x} + 
   e^A_{-x}\otimes e^B_{-x}]\label{eq:eAB+}\\
   &=\tfrac{1}{4}(1,0,0,0,0,0,0,0,1),\\ 
   	  e^{AB}_- &= \tfrac{1}{2}[e^A_{+x}\otimes e^B_{-x} + 
   e^A_{-x}\otimes e^B_{+x}]\label{eq:eAB-}\\  
   &= \tfrac{1}{4}(-1,0,0,0,0,0,0,0,1).
 \label{key_effects}
 \end{align}
Note that although these are linear combinations of product effects, namely, the four product effects $\{e^A_{+x},e^A_{-x}\}\otimes \{e^B_{+x},e^B_{-x}\}$,
they are nonetheless {\em entangled} effects in reflection-twirled boxworld because they cannot be expressed as linear combinations of {\em invariant} product effects (recall that $e^{A/B}_{+x}$ and $e^{A/B}_{- x}$ are not invariant under reflection). This is an instance of how the notion of entanglement is modified in a twirled world, a topic discussed in the main text\footnote{Furthermore, the example is of a similar form to the one introduced to demonstrate a modification of the notion of entanglement in real-amplitude quantum theory~\cite{caves2001entanglement}.}.
 Note also that these two effects sum to the unit effect, $e^{AB}_+ + e^{AB}_- = u^{AB}$, so that $\{ e^{AB}_+, e^{AB}_-\}$ describes an invariant binary-outcome measurement. As these two effects 
are the only invariant effects which are not of product form, they are 
the key to demonstrating a symmetry-induced failure of tomographic locality. That is, if one can identify a pair of invariant states on two gbits that are distinguishable {\em only} thanks to these two effects, one has a proof of the failure of tomographic locality.

 There are several such pairs of states. To generate an example, it suffices to find two invariant states whose first components (as column vectors) are distinct while they have the same value for all other components. This follows from an explicit computation of what components of an invariant bipartite state can be learned from invariant product effects. Noting that for each of the invariant effects on $A$, namely, $\mathbb{0}^A,u^A,e^A_{+y},$ and $e^A_{-y}$,
 the first component is 0, and similarly for each of the invariant effects on $B$, it follows that the first component of any invariant product effect is 0, so that one \emph{cannot} learn the first component of any bipartite state using invariant product effects.  All other nontrivial components of an invariant bipartite state, however, {\em can} be learned using invariant product effects.
The second, third, fourth and seventh components of an invariant state are necessarily 0, so the nontrivial components are the fifth, sixth and eighth. The sixth can be inferred using 
$u^A \otimes e_{+y}^B$ and $u^A \otimes e_{-y}^B$,
the eighth from 
$e_{+y}^A \otimes u^B$ and $e_{-y}^A \otimes u^B$,
and the fifth from the four product effects not involving the unit effect.

 We here provide several examples of pairs of invariant states on two gbits that are locally indistinguishable.
 
Our first example is actually a one-parameter family of pairs of states. Unlike the examples to follow, the states used here are separable from the perspective of the original boxworld (prior to twirling). 
To begin, we define two nonextremal states of a single gbit, namely, 
\begin{align}
\omega^{A}_{+} := s \omega^{A}_{+-} + (1-s)\omega^{A}_{++}\\
\omega^{A}_{-} := s \omega^{A}_{--} + (1-s)\omega^{A}_{-+},
\end{align}
where $s\in [0,1]$ is the parameter. (It is worth noting that these two states yield the same bias for the binary-outcome measurement $\{ e^A_{+y}, e^A_{-y}\}$, determined by the value of $s$.)
A parallel definition can be given for $\omega^{B}_{+}$ and $\omega^{B}_{-}$.

The pair of 
states on $AB$ are defined in terms of these as follows:
  \begin{align}
	  \omega^{AB}_+ &= \tfrac{1}{2}(\omega^{A}_+ \otimes \omega^{B}_+ + \omega^{A}_- \otimes \omega^{B}_-)\label{eq:omegaAB+}\\
   &=(1,0,0,0,(1{-}2s)^2,(1{-}2s),0,(1{-}2s),1)^T, \\
	 \omega^{AB}_- &= \tfrac{1}{2}(\omega^{A}_+ \otimes \omega^{B}_- + \omega^{A}_- \otimes \omega^{B}_+) \label{eq:omegaAB-} \\
  &= ({-}1,0,0,0,(1{-}2s)^2,(1{-}2s),0,(1{-}2s),1)^T.
 \end{align}
Above, we noted that the pair of effects in Eqs.~\eqref{eq:eAB+} and \eqref{eq:eAB-} are entangled in reflection-twirled boxworld; the same holds true for this pair of states.  That is, although each of these states is a convex mixture of product states, they are nonetheless entangled in reflection-twirled boxworld because they cannot be expressed as linear combinations of {\em invariant} product states (since $\omega^{A/B}_{+}$ and $\omega^{A/B}_{-}$ are not invariant under reflection but rather are mapped to one another). Again, this is an instance of how the notion of entanglement is modified in a twirled world.
It is straightforward to verify that $\omega^{AB}_+$ and $\omega^{AB}_-$ are {\em perfectly} distinguishable using the invariant binary-outcome measurement $\{ e^{AB}_+, e^{AB}_-\}$, but because they have the same value for all components except the first, they cannot be discriminated by any invariant product effect.

 The second and third examples make use of mixtures of different PR-box states. Note that the PR-box states themselves are not invariant.  Indeed, they can be organized into pairs constituting orbits of the $\mathbb{Z}_2$ group associated to reflection symmetry, concretely, $\{\omega^{AB}_{1},\omega^{AB}_{8}\}$, $\{\omega^{AB}_{2},\omega^{AB}_{3}\}$, $\{\omega^{AB}_{4},\omega^{AB}_{5}\}$ and $\{\omega^{AB}_{6},\omega^{AB}_{7}\}$.  Taking an equal mixture of the states in such a pair yields a state that is invariant. 
 
We now organize these invariant states into two pairs:
  \begin{equation}
 \begin{array}{cc}
	  \overline{\omega}^{AB}_{+} = \frac{1}{2}(\omega^{AB}_{2}+\omega^{AB}_{3}) =(1,0,0,0,1,0,0,0,1)^T, \\
	  \\
	 \overline{\omega}^{AB}_{-} = \frac{1}{2}(\omega^{AB}_{4}+\omega^{AB}_{5}) =(-1,0,0,0,1,0,0,0,1)^T,
 \end{array}
 \end{equation}
 and 
 \begin{equation}
 \begin{array}{cc}
	  \overline{\overline{\omega}}^{AB}_{+} = \frac{1}{2}(\omega^{AB}_{1}+\omega^{AB}_{8}) =(1,0,0,0,-1,0,0,0,1)^T \\
	  \\
	 \overline{\overline{\omega}}^{AB}_{-} = \frac{1}{2}(\omega^{AB}_{6}+\omega^{AB}_{7}) =(-1,0,0,0,-1,0,0,0,1)^T,
 \end{array}
 \end{equation}
For each of these pairs of invariant states, it is again trivial to verify that they are perfectly distinguishable using the invariant binary-outcome measurement $\{ e^{AB}_+, e^{AB}_-\}$, but because for each pair, all components except the first are the same, they are indistinguishable by invariant product effects.  

\end{document}